\documentclass{article}
\usepackage{graphicx}

\begin{document}

\centerline{\Large Origin of universes with different properties}
\bigskip

\centerline{S.G. Rubin\footnote{e-mail: sergeirubin@mtu-net.ru}}
\bigskip
\centerline{Moscow State Engineering Physics Institute, }

\centerline{Center for Cosmoparticle Physics ``Cosmion'',}

{\small The concept of a random Lagrangian is proposed. It is considered as a
basis for a new view of the old problems such as renormalization, nonzero
vacuum energy and the anthropic principle. It gives rise to nontrivial
consequences both in cosmology and in particle physics. It is shown that some
of them could be checked in the nearest future. Scalar-tensor models of gravity
are a direct and necessary consequence of this approach.}

{


\section{Introduction}

The common way of the development of physics is supported by a large amount
of observational and experimental data. On the other hand, some phenomena
which we expected to be discovered for a long time are still only
hypotheses. Other data, being very impressive, are not yet explained. As
an example, worth mentioning are the existence of dark matter and dark
energy, superhigh energy particles in cosmic rays and the ``bursts'', i.e.,
almost instantaneous energy explosions with an energy release of the order of
$10^{53}$ erg.  These astrophysical data specify the presence of new
phenomena to be comprehended.

One of the cornerstones of modern microphysics is the Weinberg-Salam
model of the electroweak interaction. Its success became evident after the
discovery of the $W^{\pm }$ and $Z$ bosons. Meanwhile, another prediction of
this model, the existence of Higgs particles, is not yet confirmed. Their
detection in modern accelerators would specify the properties of the model
and point out the direction of further development.

One could conclude that cosmology and microphysics stand before quantitative
progress which should reconcile the theoretical studies with the essentially
new experimental data. It makes sense to analyze the basic postulates which
underlie every modern theory.

A widespread approach consists in choosing dynamical variables and a form of
a Lagrangian from the beginning. It is supposed that the smaller is the
number of parameters the better is the quality of a theory. This approach,
being successful in areas of specific calculations, suffers from some
problems.  These problems are rarely discussed and usually are even not
mentioned. Indeed, let us suppose for a minute that somebody succeeded in his
attempts to create a theory explaining all experimental data. It does not
mean the end of physics. On the contrary, it becomes absolutely clear
that several general questions should be resolved:

(A) What mechanism is responsible for the specific form of a potential with
particular values of the parameters? In particular, why do we suppose that
the minimum of the potential is exactly zero?

(B) To which extent the quantum corrections deforming the shape of
the potential are important?

(C) The question of renormalizability of a theory appears not to be so
simple if one takes into account the interaction of particles with gravitons.
The latter is usually extremely small and surely should not be considered in
real calculations, but its existence is of principal importance. Gravity,
being a non-renormalizable theory, leads to the same property of any theory
connected with it.

These problems are not very important for low-energy physics, and the
scientists usually wave them away. But modern accelerators have reached rather
high energies of the order of $1 $ TeV and higher. Not only the
experimentalists, but the theorists as well feel the necessity of dealing with
a new level of energies: theories of the early Universe operate with Planck
densities.

Problem (A) became topical after the discovery of dark energy \cite{Riess98},
which could be explained most easily as a nonzero vacuum energy density. The
latter, being $\sim 120$ orders of magnitude smaller than Planck scale, allows
the formation of the large-scale structure of our Universe.  This problem, or
more widely, the problem of creation of the Universe with the observable
properties, has attracted attention of many scientists and caused a prolonged
discussion \cite{Carr79, Rozental80e}. This discussion continued until recently
\cite{Star99, Banks00, Wein00, Ru39, Dolgov02}.

The anthropic principle plays a significant role in the discussion of the
problem of fine tuning of the Universe parameters and, in particular, the
problem of a nonzero dark energy \cite{Banks00, Wein00, Vil00a, Ru36, Ellis03}.
In general, this principle suggests the existence of a set of universes with
different properties. Some of them are similar to our Universe. The others, in
fact, a dominant majority, are not suitable for life formation. ``The only
thing that remains'' is to create a theory which could base the existence of
such a set of universes. In this paper we argue that modern quantum field
theory supplies us with the necessary ingredients for solving this problem.

Neglecting quantum corrections (problem (B)) seems doubtful at high energies.
Moreover, new terms in the Lagrangian appearing due to quantum corrections
are used creation of inflationary models \cite{Linde90} and particle models
\cite{Wudka01}. It is evident that quantum corrections carry on a double
meaning. On the one hand, they lead to a significant uncertainty of
predictions of any inflationary model. On the other hand, the same
corrections give new possibilities for particle models and hence for models
of the early Universe tightly connected with them.

A new approach is proposed in this paper. It is supposed \textit{a priori}
that the contribution of quantum corrections to a Lagrangian must not be
neglected. This supposition permits one to validate the existence of
universes with different properties and look from another side at the
problems listed above. It is shown that an experimental test of the
suggested approach is possible on the basis of modern accelerators.

\section{Deformation of a potential}

According to the previous discussion, any fixing of the shape of a
potential leads inevitably to problems (A)--(C) listed in the Introduction.
On the other hand, our Universe has been formed at very high energies where
the contribution of the quantum corrections was important. Evidently, the
potential acquires a much more complex form, compared with the original
choice, due to quantum corrections. If we restrict ourselves to scalar
fields $\varphi$, naive calculations of the quantum corrections lead to a
potential in the form of an infinite set
\begin{eqnarray} 
U_{qc}(\varphi ) = \sum\limits_{k}^\infty {a_k \varphi ^k }.  \nonumber
\end{eqnarray}
Generally speaking, negative powers are not excluded. Unfortunately,
calculation of the coefficients $a_k$ is waste effort for two reasons.
Firstly, it is hard to believe that this decomposition is correct at large
values of the field $\varphi$. Secondly, each term of this sum is a result
of superposition of interactions with particles of every sort. Their
contributions vary unexpectedly with increased degree of a term.
Consequently, any information about the shape of the potential in the
vicinity of a chosen field value $\varphi_0$ is useless at $\varphi \gg
\varphi_0$.

Thus any model of elementary particles with a specific form of the potential
postulated from the beginning with a small number of parameters is doomed to
failure at large values of the dynamical variables. As a possible solution of
the problem, a new postulate is proposed. The latter is an analogue of the
concept of attenuation of correlations known in statistical physics. As in
the latter, the probabilistic approach used below allows one to obtain new
results and to clarify the already known problems.

Let us first of all introduce the concept of probability density $P(V;\varphi
)$ of finding a specific value $V$ of the potential at a given field value
$\varphi$.  It is supposed that the functions $V(\varphi )$ and $\varphi
(x,t)$ are smooth enough except a discrete set of points. Then the only
requirement to the form of the potential is expressed as the following
postulate:

\begin{flushleft} {($\ast$){\it \hspace{0.314 cm}Let the value of the potential $V_0$ be known at a
given field value $\varphi_0$. Then there exists such a value $\Phi$ ($0<\Phi
<\infty$) that, for any $V$ and $\varphi$, $P (V;\varphi )>0$ provided
$|\varphi - \varphi_0|>\Phi$.}}\end{flushleft}

The value $\Phi$ depends on specific values of the potential $V_0$ and the
dynamical variable $\varphi_0$. As we shall see later, the only important
thing is its finiteness, as is mentioned in the postulate. The transition
to a probabilistic description is a cornerstone of this approach. It makes
our possibilities much richer, as has happened in the transition from the
classical description to the quantum-mechanical one. It is worth emphasizing
that the form of the potential is not postulated from the beginning.
Instead, only one, rather general property of the potential is proposed.
Nevertheless, this property leads to multiple consequences which may could
be experimentally tested. As we will show below, some of them could be
performed in the nearest future, validating or invalidating the postulate.

Let us now discuss the general corollaries of the postulate ($\ast$). The
first direct corollary is that such a potential possesses a countable set of
zeros.  Indeed, if we make the inverse supposition, then, starting from some
field value $\overline{\varphi}$, the function $V(\varphi )$ is only
positive or only negative for $|\varphi |>\overline{\varphi}$. Consequently,
$P(V<0)=0$, or $P(V>0)=0$ in this case, which contradicts the postulate
($\ast$). It is obvious that a countable set of zeros implies a countable
set of extrema of the potential.

To proceed, let one know that some minimum of the potential takes place at
the field value $\varphi =\varphi_m$. Then, according to postulate
($\ast$), there is a probability $P(V_m )dV_m > 0$ of finding the potential
value in a given interval $(V_m ,V_m +dV_m )$. It immediately follows that
there exists a countable set of such minima in the interval in question.
The figure shows a representative form of the scalar field potential within
some interval. If one considers the scalar field as an inflaton, the
universe formation takes place at the minima $m-1,m,m+1,m+2...$. Hence, a
transparent consequence of the postulate ($\ast$) is prediction of a nonzero
cosmological constant because the probability to find a local minimum
with a preset energy density is zero.

\begin{figure}[tbp]
\hspace{0cm}
\includegraphics[width=16cm,
height=8cm]{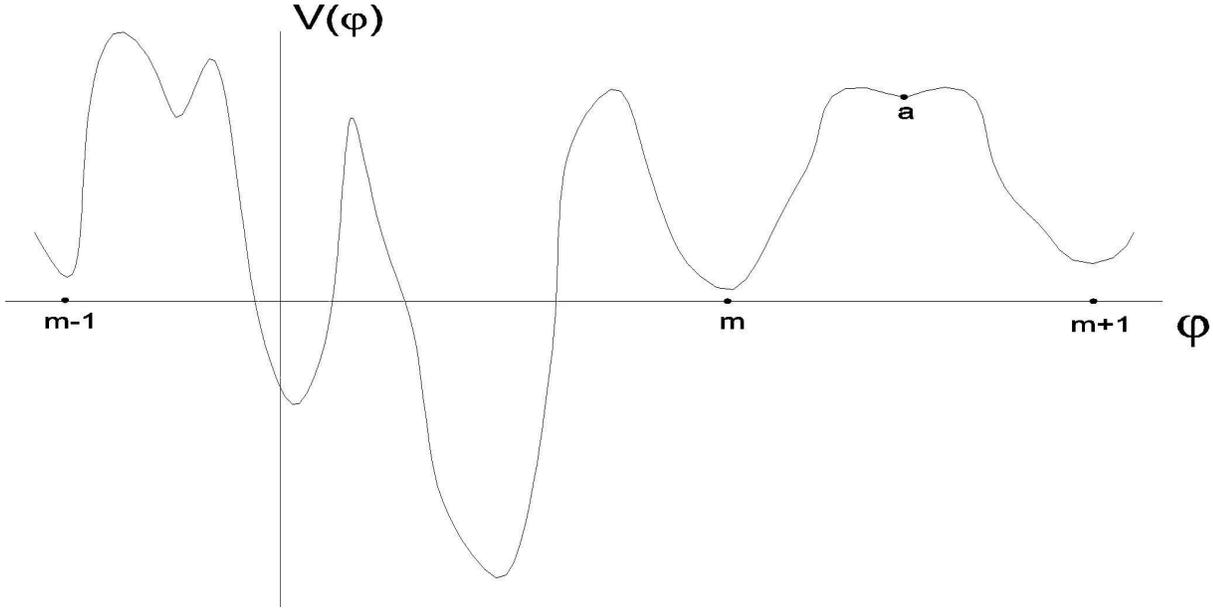} \caption{ Characteristic form of the potential
discussed in the text. Points mark minima where universes of different sorts
are formed and evolve. Causal connection between the universes is absent.}
\label{randpot}
\end{figure}

The shape of this ``random'' potential is unique in the vicinity of each
minimum and hence the process of inflation is unique as well. Minimum values of
the potential (vacuum energy density) are materialized with some probability
density which is nonzero for any values. Hence, there is a countable set of
low-lying minima, which is a necessary, but not sufficient, condition for the
formation of universes similar to our Universe.

The most unpleasant point is the occurrence of unstable areas with $V < 0$.
A similar situation was revealed and discussed earlier. For example, the
quantum corrections from interaction with fermions can result in scalar
field potentials unbounded from below (see, e.g., \cite{Krive76}). An
accurate renormalization of a Higgs-like potential leads to new minima
\cite{Chang75}, some of them being unstable. However, the spatial areas with
$V < 0$ are not causally connected with the visible part of our Universe. It
is a result of an initial inflationary period of evolution of the Universe
when its size has increased up to $\sim \exp (10 ^ {12}) $ cm  in the modern
epoch, many orders of magnitude greater than the size of the observable
universe $\sim 10 ^ {28} $ cm.

The potential can be approximated by the first terms of a Taylor series. It
can be easily written in the vicinity of one of the minima, say,
$\varphi_m$:
\begin{equation}
V(\varphi ) = V(\varphi _m + \phi ) \simeq V(\varphi _m ) + a\phi ^2 + b\phi
^4 ,
\end{equation}
thus simulating various potentials, which are usually postulated from the
beginning.

A logical extension of the previous discussion is the inclusion of matter and
gauge fields. It needs a generalization of postulate $(\ast)$ to any
parameters of the theory, which are deformed by quantum corrections.

\begin{flushleft}{\it $(\ast \ast )$\hspace{0.314cm} All quantities of the theory, which are deformed by quantum corrections,
comply with postulate $(\ast )$.}
\end{flushleft}

In a substantiation of this postulate it is possible to refer to the same
reasoning that has resulted in postulate ($\ast $). Here it is important to
note, that, due to the quantum corrections, all parameters of the theory $%
g_n,\ n=1,2... N$, \ ($N $ is the number of parameters of the theory) turn
into (random) functions of a scalar field $g_n (\varphi)$. The values of a
scalar field $\varphi_m $ deliver minima of the potential, but not of the
functions $g_n (\varphi)$. The values $g_n (\varphi_m )$ are considered as
constants in ordinary low-energy physics.

\section{Weinberg--Salam model}

Let us consider modifications of the $SU(2)\bigotimes U(1)$ Weinberg--Salam
Standard Model (SM), caused by the postulates $(\ast )$ and $(\ast\ast )$.

\subsection{Higgs field}

The potential of the Higgs field $\chi = {\chi_1 \choose \chi_2}$
is usually supposed in the form
\begin{equation}  \label{PotHiggs}
V_{Higgs} (\chi ) =\frac{\lambda}{4} \left( {\left| \chi \right|^2 - v^2 /2 }
\right)^2 .
\end{equation}
i.e., a nonzero vacuum average is postulated from the beginning. A pleasant
gift of our postulates is that it is now unnecessary --- quantum
corrections lead to multiple minima of any potential, including the Higgs
one. The problem is more complex because the Higgs field necessarily
interacts with the inflaton (at least due to multiloop corrections), and we
inevitably obtain a set with an infinite number of terms,
\begin{equation}  \label{IPH}
V(\varphi ,\chi ) = \sum\limits_{k,n=0}^\infty {a_{kn} \varphi ^k |\chi|
^{2n}}
\end{equation}

Application of postulate ($\ast $) results in a potential of a complex form
with a lot of extrema and valleys in various directions. Minima of the
potential are situated at points with the field value $(\varphi_m , v_m )$,
$v_m \equiv \sqrt{|\chi_m |^2}$. One can see that there is no need to include a
nonzero vacuum average artificially. Of course, the probability that two
neighboring vacua appear to be symmetrical, as it takes place in the SM, is
negligible. It should not disturb us for this symmetry is not an important
feature of the model. In this connection, we would like to mention
Ref.\,\cite{Froggatt01} devoted to the Weinberg--Salam model with two
asymmetric vacua. One of them is located at the Planck scale.

\subsection{Modernization of the Standard Model}

The Weinberg--Salam model is described by the Lagrangian
\begin{eqnarray}
&&L=L_{gauge}+L_{lept}+L_{scalar}+L_{int}; \nonumber \\
&&L_{gauge}=-\frac{1}{4}F_{\mu \nu }^{i}F^{i\mu \nu }-\frac{1}{4}B_{\mu \nu
}B^{\mu \nu };  \nonumber \\
 &&L_{lept}=\bar{R}i\left(
{\widehat{\partial}+ig^{\prime }\widehat{B}}\right) R+  \nonumber \\
 &&\bar{L}i[\widehat{\partial}+i\frac{g^{\prime }}{2}\widehat{B}-i\frac{g}{2}\tau
^{i}\widehat{A^{i}}]L;  \nonumber \\ &&L_{scalar}=\left\vert
{[\widehat{\partial}-i\frac{g^{\prime }}{2}\widehat{B}-i\frac{g }{2}\tau
^{i}\widehat{A^{i}}]\chi }\right\vert ^{2}-V(\left\vert \chi \right\vert ^{2});
\nonumber
\\ &&L_{{int}}=-G\left[ {\bar{R}\chi ^{+}L+\bar{L}\chi R}\right] , \label{VS}
\end{eqnarray}
with the notations
\begin{eqnarray}
&&F_{\mu \nu }^{i}=\partial _{\mu }A_{\nu }^{i}-\partial _{\nu }A_{\mu
}^{i}+g\varepsilon ^{ijk}A_{\mu }^{j}A_{\nu }^{k}, \nonumber\\  &&B_{\mu \nu
}=\partial _{\mu }B_{\nu }-\partial _{\nu }B_{\mu }.  \nonumber\\
&&V(\left\vert \chi \right\vert ^{2})=\lambda (\left\vert \chi \right\vert
^{2}-v^{2}/2). \label{Vsdenot}
\end{eqnarray}
Here $A$ and $B$ are the gauge fields, $L$ and $R$ are the left and right
fermions.  The Higgs field $\chi$ has the nonzero vacuum average ${0\choose
v/\sqrt{2}}$. The parameters $g,\ g',\ v,\ G$ of the model are expressed in
terms of observable values: mass of the charged $W$ boson, $M_{W}$, mass of
the $Z$ boson, $M_{Z}$, the electron charge $e$ and the Fermi constant
$G_{F}$:
\begin{eqnarray}
&&M_W = gv/2,\quad M_Z = \frac{v}{2}\sqrt {g^2 + g^{\prime 2} } , \nonumber\\
&&e= \frac{{gg^{\prime}}}{{\sqrt {g^2 + g^{\prime 2} } }}, \quad G_F =
\frac{1}{\sqrt 2 v^2 }.  \label{L8}
\end{eqnarray}

As was shown in Ref.\,\cite{Taylor71}, the renormalization procedure can
be chosen in such a manner as to preserve the gauge invariance of the
Lagrangian (\ref{VS}). That is why we restrict ourselves to studying the
results of renormalization of the parameters $g,\ g',\ G$. According to
postulates $(\ast )$ and $(\ast\ast )$, these parameters transform into
random functions:
\begin{equation}  \label{L9}
g,g^{\prime},G\Rightarrow g(\varphi,|\chi |^2),g^{\prime}(\varphi,|\chi
|^2),G(\varphi,|\chi |^2 ).
\end{equation}

The proof is just the same as for the random potential.

At the same time, the potential of the Higgs field $V(|\chi |^2 )$ acquires
the form (\ref{IPH}) due to the same postulates $(\ast )$, $(\ast\ast )$.

Let us now discuss the last term in the expressions (\ref{VS}) describing
fermion interaction with Higgs particles. It will be shown that it gives a
nontrivial result, different from the SM prediction and capable of an
experimental test. Replacement of the constants $g,\ g',\ v,\ G$ with the
functions (\ref{L9}) leaves a trace in an effective Lagrangian.

The expression $L_{\rm int}$ in the Standard Model permits one to obtain the
fermion mass $m_f$ uniquely related to the vacuum average $v/\sqrt{2}$
inserted by hand,
\begin{equation}  \label{L10}
m_f = G v/\sqrt{2}.
\end{equation}
If one takes into account that the parameter $G$ is a random function of the
field $\varphi$, the result appears to be rather different. Let us parameterize
the Higgs doublet as usual: $\chi = e^{i\Theta (x) }{0\choose \chi ^0 }$ where
$\Theta (x)$ is an $SU(2)$ matrix to be removed from the final Lagrangian by a
gauge transformation. As a result, the term in question is
\begin{equation}  \label{L12}
L_{int}=-G(\varphi ,(\chi^0 ) ^2 )\chi^0 \overline{f_R }f_L .
\end{equation}

Let a minimum No.\,$m$ of the potential (\ref{IPH}) take place at the field
value ($\varphi_m$, $\chi^0_m$) $(\chi^0_m =v/\sqrt{2}$ in the usual
notations).  The first terms in the Taylor expansion of Eq. (\ref{L12}) in
powers of $h = \chi^0 -\chi_m ^0$ have the form
\begin{eqnarray}
&&L_{int}\simeq  - \chi_m ^0\cdot G(\varphi_m ,(\chi_m ^0 ) ^2 ) \overline{f_R
}f_L -  \nonumber \\
 &&G_f h\overline{f_R }f_L ,  \nonumber \\
&&G_f= \frac{\partial G(\varphi_m ,(\chi_m ^0 ) ^2 )} {\partial \chi_m ^0}%
\chi_m ^0 + G(\varphi_m ,(\chi_m ^0 ) ^2 ).\label{L13}
\end{eqnarray}
The fermion mass
\begin{equation}  \label{L12a}
m_f =\chi_m ^0\cdot G(\varphi_m ,(\chi_m ^0 ) ^2 )
\end{equation}
appears to depend on the number $m$ of the specific minimum (see also
\cite{Donoghue97}), in which the universe evolves.

The field values $\varphi_m ,\ \chi^0_m$ represent a minimum of the
potential (\ref{IPH}), but not the minima of the other functions like
$G(\varphi_m ,\ (\chi_m ^0 ) ^2 )$. That is why the usual proportionality of
the fermion mass $m_f$ and its coupling constant with the Higgs particle is
absolutely absent.

Another deviation from the SM could be found in the self-interaction couplings
of the Higgs particles. For example, the constant of trilinear interaction in
the framework of the Weinberg--Salam model is equal to $\lambda_{hhh} =
3\sqrt{2}\lambda v$ in the zeroth order of perturbation theory and hence is
proportional to the known vacuum average value $v$. A different prediction
follows from the potential (\ref{IPH}). This constant has the form
$\lambda_{hhh} = (1/6)\partial^3 V(\varphi_m,\ \chi_m )/\partial \chi_m ^3$ and
has no connection with the other parameters.

It becomes evident that this approach is able to restore the Weinberg--Salam
model with the exception of interactions with the Higgs particles $h$. It is
instructive to consider the part of the Lagrangian $L_{\rm lept}$
responsible for the interaction of leptons (electrons) with the gauge
fields. After the standard replacement of the fields $B_{\mu},\ A_{\mu}^i$
with the physical fields $W_{\mu}^{\pm},\ Z_{\mu},A_{\mu}$ one obtains
\cite{Itzykson}
\begin{eqnarray}
&&L_{ lept} = C_1 \left[ {\bar \nu _e \gamma ^\mu (1 - \gamma _5 )e W_\mu ^ + +
h.c.} \right] -  \nonumber \\  &&- C_2 \left( {2\bar e_R \gamma ^\mu e_R + \bar
\nu _e \gamma ^\mu \nu _e + \bar e_L \gamma ^\mu e_L } \right)Z_\mu +
\nonumber \\ &&+ C_3 \left( {\bar e_L \gamma ^\mu e_L - \bar \nu _e \gamma ^\mu
\nu _e } \right)Z_\mu - eA_\mu \bar e\gamma ^\mu e . \label{L14}
\end{eqnarray}

The quantities $C_{i},\ e$ are expressed in a usual way in terms of the
initial parameters
\begin{eqnarray}  &&C_{1}=\frac{g}{{2\sqrt{2}}};\quad
C_{2}=\frac{e}{2}\tan\theta _{W};\quad C_{3}=\frac{e}{2}\cot\theta _{W} ,
\nonumber \\ &&e=\frac{{gg^{\prime }}}{{\sqrt{g^{2}+g^{\prime }{}^{2}}}};\quad
\tan\theta _{W}=g^{\prime }/g  \label{L15}
\end{eqnarray}
and, according to (\ref{L9}), are functions of the fields $\varphi $ and
$\chi$. Both fields are located in the vicinity of some minimum of the
potential (\ref{IPH}), and we can restrict ourselves to the first terms in the
Taylor expansion:
\begin{eqnarray}
&&C_{i}\simeq C_{i}(\varphi _{m},\chi _{m}^{0})+\frac{{\partial C_{i}(\varphi
_{m},\chi _{m}^{0})}}{{\partial \chi _{m}^{0}}}h;  \nonumber \\ &&e\simeq
e(\varphi _{m},\chi _{m}^{0})+\frac{{\partial e(\varphi _{m},\chi
_{m}^{0})}}{{\partial \chi _{m}^{0}}}h. \label{L16}
\end{eqnarray}

The inflaton field is supposed to be rather massive, so that we can neglect any
interaction with its quanta. Indeed, its value $\sim 10^{13}$ GeV
\cite{Linde91} is many orders of magnitude greater than the masses of ordinary
particles. The inflaton dynamics is important only at a very early stage of the
Universe evolution, when the horizon size is smaller than the inverse inflaton
mass $1/m_{\rm inflaton}$. The first terms in the expansion (\ref{L16}) were
determined experimentally as far as they are connected with the known
parameters: the electron charge $e$ and the Weinberg angle $\theta_W$ according
to Eqs. (\ref{L15}).  The second terms are responsible for new vertices of
interaction of the leptons with the Higgs field quanta $h$. More certainly, the
vertices $\bar \nu_e Weh,\ \bar{e} eZh,\ \bar \nu_e \nu_e Zh,\ \bar{e} eAh$
originate from the Lagrangian (\ref{L14}) and Eqs (\ref{L15}), (\ref{L16}) in
the form
\begin{eqnarray}
&&L^{\prime}_{ lept} = \Gamma_1 \left[ {h\bar \nu _e \gamma ^\mu (1 - \gamma _5
)e W_\mu ^ + + h.c.} \right] -  \nonumber \\ &&- \Gamma_2 \left( {2h\bar e_R
\gamma ^\mu e_R + h\bar \nu _e \gamma ^\mu \nu _e + h\bar e_L \gamma ^\mu e_L }
\right)Z_\mu + \nonumber  \\ &&+ \Gamma_3 \left( {h\bar e_L \gamma ^\mu e_L -
h\bar \nu _e \gamma ^\mu \nu _e } \right)Z_\mu - \Gamma_e hA_\mu \bar e\gamma
^\mu e . \label{L17}
\end{eqnarray}

The four coupling constants $\Gamma_1$, $\Gamma_2$, $\Gamma_3$ and $\Gamma_e$
are expressed in terms of two unknown parameters, $ B\equiv \partial g/\partial
\chi_m^0$ and $B' \equiv\partial g' /\partial \chi_m^0$:

\begin{eqnarray}
&&\Gamma _e  = \left( {\frac{e}{{g'}}} \right)^3 B + \left( {\frac{e}{g}}
\right)^3 B', \nonumber  \\ &&\Gamma _1  = \frac{1}{{2\sqrt 2 }}B, \label{L18a}
\\ &&\Gamma _2  = \frac{1}{2}\frac{{eg'}}{{g^2 }}\left( {\frac{{e^2 g}}{{g'^3
}} - 1} \right)B + \frac{1}{2}\frac{e}{g}\left( {\frac{{e^2 g'}}{{g^3 }} + 1}
\right)B',  \nonumber   \\ &&\Gamma _3  = \frac{1}{2}\frac{{eg}}{{g'^2 }}\left(
{\frac{{e^2 g'}}{{g^3 }} - 1} \right)B' + \frac{1}{2}\frac{e}{{g'}}\left(
{\frac{{e^2 g}}{{g'^2 }} + 1} \right)B . \nonumber
\end{eqnarray}
These coupling constants could be measured independently. On the other hand,
they are connected with each other, as readily follows from (\ref{L18a}):
\begin{eqnarray}
&&\Gamma _2 = \Gamma _e \frac{1}{2}\left( {\frac{g}{e}} \right)^2 \left(
{\frac{{e^2 g^{\prime}}}{{g^3 }} + 1} \right)  +  \nonumber \\ && \Gamma_1
\sqrt 2 \left[ {\frac{{eg^{\prime}}}{{g^2 }}\left( {\frac{{e^2 g} }{{g^{\prime
3 }}} - 1} \right) - \frac{e}{g}\left( {\frac{g}{{g^{\prime}}}} \right)^3
\left( {\frac{{e^2 g^{\prime}}}{{g^3 }} + 1} \right)} \right] , \nonumber \\
&&\Gamma _3 = \Gamma _e\frac{1}{2}\left( {\frac{g}{e}} \right)^3 \frac{{eg
}}{{g^{\prime 2} }}\left( {\frac{{e^2 g^{\prime}}}{{g^3 }} - 1} \right)  +
\label{L18}\\ && \Gamma _1\sqrt 2 \left[ {\frac{e}{{g^{\prime}}}\left(
{\frac{{e^2 g}}{{ g^{\prime 3 } }} + 1} \right) - \frac{{eg}}{{g^{\prime 2}
}}\left( {\frac{g}{{ g^{\prime}}}} \right)^3 \left( {\frac{{e^2
g^{\prime}}}{{g^3 }} - 1} \right)} \right] .   \nonumber
\end{eqnarray}

These expressions give a theoretical prediction for the values of $\Gamma_2$
and $\Gamma_3$ provided the vertices  $\Gamma_e$ and  $\Gamma_1$ are
determined experimentally. At the same time, the same vertices $\Gamma_2$
and $\Gamma_3$ could be measured independently for comparison with the
theoretical result. The existence of new vertexes (\ref{L17}) obeying the
relations (\ref{L18}) is a direct consequence of the initial postulates
$(\ast )$ and $(\ast \ast )$. On the other hand, they may be checked
experimentally in the nearest future. For example, the properties of the
Higgs bosons will be investigated in the modern accelerators. Such a
possibility is discussed in Ref.\,\cite{Castanier01} for $e^+ e^-$
annihilation at energies $\sim 500$ GeV in the center of mass system. In
case the Higgs bosons is discovered, it will allows us to validate or
invalidate the relations (\ref{L18}) and hence the above postulates.

\section{Discussion}

In this paper, the new postulates $(\ast )$ and $(\ast \ast)$ have been
proposed. These postulates are used instead of strictly fixing the
Lagrangian from the beginning. This concept gives rise to nontrivial
consequences both in cosmology and in particle physics. First of all, it
allow one to prove the existence of a countable set of universes disposed in
potential minima with different values of microscopic parameters. It serves
as a necessary ingredient of the anthropic principle which permits one to
explain the origin of a universe like ours.

The problems listed in the Introduction look quite solvable if one uses the
probability language for formulation of the postulates. In this framework,
the answer to problem (A) is: ``there exist an infinite set of
universes with different parameters in the vicinity of local minima. One
could extract from this set of universes a subset with parameters close to
those of our Universe. The minimal value of the potential is not zero.
Meanwhile, there exists an infinite set of universes with sufficiently
small values of the potential minimum. It is a necessary condition for
nucleation and formation of a universe like ours''.

An additional pleasant property of the potential governed by the postulates
is its absolute renormalizability (problem (C)). Indeed, it contains terms
with any power of the scalar field from the beginning, and any quantum
correction can be included in the Lagrangian parameters.

The answer to the question (B) is evident. It is quantum corrections that
deform the potential and produce an infinite set of minima.

Postulates $(\ast )$ and $(\ast \ast)$ necessarily lead to scalar-tensor
theories of gravity. Indeed, if we wish to be consistent, we have to admit,
according to postulates $(\ast )$, $(\ast \ast)$, that the quantum
corrections convert {\it all\/} parameters of a Lagrangian, including the
gravitational constant $G_N$, into random functions. In this case, a general
form of the Lagrangian can be readily written:
\begin{equation}  \label{L22}
L = - \frac{F(\varphi )}{{16\pi G_N }}R + \frac{K(\varphi )}{2}\left( {%
\partial \varphi } \right)^2 - V_{ren}(\varphi ).
\end{equation}

An immediate conclusion from Eq. (\ref{L22}) is that Newton's gravitational
constant is only an effective one and varies in different universes enumerated
by the number $m$: $G_N(m) = G_{N, {\rm our}} F(\varphi_m ) /F(\varphi_{\rm
our} )$ (the index `our' relates to our Universe). Different sorts of such
theories are discussed in, e.g., Refs.\,\cite{Bergmann68, Fradkin85, Torres97}.
Their global properties are considered in Ref.\,\cite{Bronnikov02}.

An important thing is that, in spite of the generality and brevity of the
postulates, they possess a predictive power, and some of their predictions
may be tested in the nearest future. More definitely,

(a) new vertices of interactions of leptons, gauge fields and Higgs bosons
appear in this framework. One-to-one connections (\ref{L18}) between
them may be checked experimentally in the nearest future;

(b) a strict connection between fermion masses and the coupling constant of
their interaction with scalar particles is absent. It could be very
important for axion models because it strongly facilitates the limits
following from the observational data \cite{Khlopov};

(c) the trilinear interaction constant appears to be a free parameter,
contrary to the prediction of the Weinberg--Salam Standard Model;

(d) the cosmological $\Lambda$-term is not zero. More definitely, $\Lambda =
Const$, contrary to the quintessence models.

A serious argument against the proposed postulates would be strict a equality
to zero of the cosmological constant because such universes have measure
zero among others. The evidence that the cosmological constant is not
zero is rather firm these days, but we still cannot exclude the opposite
possibility. If the future observations indicate that, nevertheless,
$\Lambda =0$, it will make the postulates doubtful.

The author is grateful to A.V. Berkov, E.D. Zhizhin, A.Yu. Kamenschik, V.M.
Maximov, A.S. Sakharov and M.Yu. Khlopov for their discussions. This work is
partly performed in the framework of State Contract 40.022.1.1.1106 and
supported in part by RFBR grant 02-02-17490 and grant UR.02.01.026.


\begin{thebibliography}{10}
\bibitem{Riess98}
A.G.Riess, Astron. Journal. {\bf 116},  1009  (1998).

\bibitem{Carr79}
B.J. Carr and M.J. Rees, Nature {\bf 278},  605  (1979).

\bibitem{Rozental80e}
I.L. Rosental, UFN {\bf 131},  239  (1980).

\bibitem{Star99}
V.Sahni and A.Starobinsky, Int.J.Mod.Phys. {\bf D9},  373  (2000).

\bibitem{Banks00}
T. Banks, M. Dine, and L. Motl, JHEP  031  (2001).

\bibitem{Wein00}
S.Weinberg, astro-ph/0005265  (2000).

\bibitem{Ru39}
S.G. Rubin, Chaos, Solitons and Fractals (CHAOS2013) {\bf 14},  891  (2002).

\bibitem{Dolgov02} A. Dolgov, hep-ph/0203245.

\bibitem{Vil00a}
J.Garriga and A.Vilenkin, Phys.Rev. {\bf D64},    (2001).

\bibitem{Ru36}
S.G.Rubin, Gravitation \& Cosmology, Supplement {\bf 8},  53  (2002).

\bibitem{Ellis03} G.F.R. Ellis, U. Kirchner and W.R. Stoeger, astro-ph/0305292.

\bibitem{Linde90}
A.~D. Linde, {\em The Large-scale Structure of the Universe} (Harwood Academic
  Publishers, London, 1990).

\bibitem{Wudka01}
J. Wudka and B. Grzadkowski, JHEP,PRHEP-hep2001  147  (2001).

\bibitem{Krive76}
I.V. Krive and A.D. Linde, Nucl. Phys. {\bf B117},  265  (1976).

\bibitem{Chang75}
S.-J. Chang, Phys.Rev. {\bf D12},  1071  (1975).

\bibitem{Froggatt01}
C.D. Froggatt, H.B. Nielsen, and Y. Takanishi, Phys.Rev. {\bf D64},  113014
  (2001).

\bibitem{Taylor71}
J.C. Taylor, Nucl. Phys. {\bf B33},  436  (1971).

\bibitem{Donoghue97}
J.F.Donoghue, Phys.Rev. {\bf D57},  5499  (98).

\bibitem{Itzykson}
C. Itzykson and J.-B. Zuber, {\em Quantum Field Theory} (McGraw-Hill, New York,
  1984).

\bibitem{Linde91}
A.~D. Linde, Physica Scripta {\bf T36},  35  (1991).

\bibitem{Castanier01}
C. Castanier, P. Gay, P. Lutz, and J. Orloff, hep-ex/0101028  .

\bibitem{Bergmann68}
P.G. Bergmann, Int.J.Theor.Phys. {\bf 1},  25  (1968).

\bibitem{Fradkin85}
E.S. Fradkin and A.A. Tseytlin, Phys.Lett. {\bf B158},  316  (1985).

\bibitem{Torres97}
D. Torres and H. Vucetich, Phys.Rev. {\bf D54},  7373  (1996).

\bibitem{Bronnikov02}
K.A. Bronnikov, J. Math. Phys. {\bf 43}, 6096 (2002), gr-qc/0204001.

\bibitem{Khlopov}
M.Yu. Khlopov, {\em Cosmoparticle Physics} (World Scientific, Singapore-New
  Jersey-London-Hong Kong, 1999).

\end{thebibliography}
\end{document}